# YOLoC: DeploY Large-Scale Neural Network by ROM-based Computing-in-Memory using ResiduaL Branch on a Chip

Yiming Chen, Guodong Yin, Zhanhong Tan, Mingyen Lee, Zekun Yang, Yongpan Liu, Huazhong Yang, Kaisheng Ma and Xueqing Li
BNRist/ICFC, Electronic Engineering Department, Tsinghua University, Beijing, China
Email: xueqingli@tsinghua.edu.cn

## ABSTRACT

Computing-in-memory (CiM) is a promising technique to achieve high energy efficiency in data-intensive matrix-vector multiplication (MVM) by relieving the memory bottleneck. Unfortunately, due to the limited SRAM capacity, existing SRAM-based CiM needs to reload the weights from DRAM in large-scale networks. This undesired fact weakens the energy efficiency significantly. This work, for the first time, proposes the concept, design, and optimization of computing-in-ROM to achieve much higher on-chip memory capacity, and thus less DRAM access and lower energy consumption. Furthermore, to support different computing scenarios with varying weights, a weight fine-tune technique, namely Residual Branch (ReBranch), is also proposed. ReBranch combines ROM-CiM and assisting SRAM-CiM to achieve high versatility. YOLoC, a ReBranch-assisted ROM-CiM framework for object detection is presented and evaluated. With the same area in 28nm CMOS, YOLoC for several datasets has shown significant energy efficiency improvement by 14.8x for YOLO (DarkNet-19) and 4.8x for ResNet-18, with <8% latency overhead and almost no mean average precision (mAP) loss (-0.5% ~ +0.2%), compared with the fully SRAM-based CiM.

## CCS CONCEPTS

• Hardware ~ Integrated circuits ~ Semiconductor memory ~ Read-only memory • Computing methodologies ~ Artificial intelligence ~ Computer vision ~ Computer vision problems ~ Object detection

## KEYWORDS

Computing-in-Memory, ROM-CiM, Read-Only Memory, YOLoC.

## 1 Introduction

Convolutional neural network (CNN) has been widely used in computer vision and other applications. To support more scenarios, generally, increasing parameters in CNN are being used in complex models. For example, Tiny-YOLO and YOLO have 11.3 M and 46 M weights, respectively. In a custom accelerator chip design, to reduce the memory-wall overheads of energy and latency in frequent DRAM data accesses [1][2], computing-in-memory (CiM) techniques have been proposed based on CMOS SRAM/eDRAM[3][4][5] and other beyond-CMOS technologies [6][7]. Among these existing techniques, SRAM-based CiM is intriguing because of its high design flexibility and mature fabrication support. Unfortunately, SRAM and SRAM-based CiM face the challenge of low density: it is challenging to store all weights inside one chip for large neural networks. Consequently, the data in the SRAM need to be dumped and reloaded from the DRAM, which results in undesired energy and latency costs [8]. Notice that, although reusing the weight data reduces the total times of DRAM access [9], loading the weights from DRAM is a bottleneck towards high energy efficiency [10].

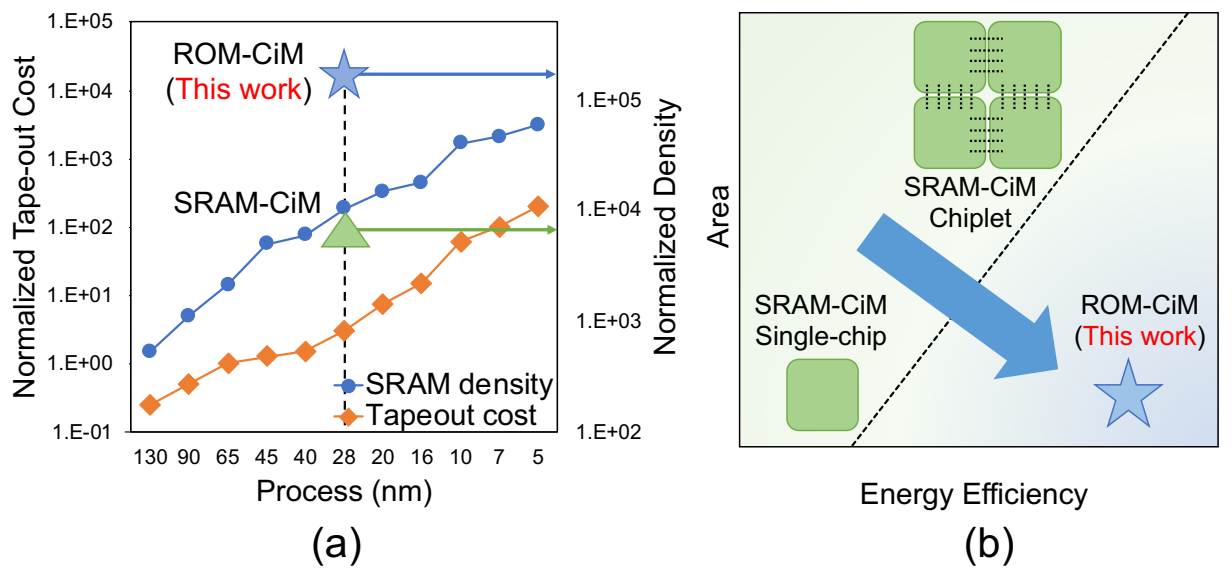

**Figure 1: Comparisons between methods to store more weights, (a) technology scaling, (b) chiplet integration.**

One approach to the mitigation of the memory capacity dilemma is using an advanced process technology with scaled-down transistors. As shown in Figure 1(a), apparently, this approach could be difficult due to the soaring fabrication costs at smaller technology nodes. This fact leads to a fundamental question: *why don't we use the much denser ROM for CiM?*

A seemingly reasonable "answer" could be the limitation of the essentially read-only capability of ROM, which makes it impossible to update the weights stored in ROM. Differently, this paper tells a new story with highlighted contributions:

First, by using the much denser ROM, we show that it is practical to store all the weights of relatively large networks, e.g. YOLO (Darknet-19 backbone), in a single cm-scale CMOS chip at 28nm. This makes it possible to dramatically reduce and even prevent off-chip DRAM weight access for higher energy efficiency. In addition, ROM is essentially non-volatile, leading to standby power savings. To showcase this opportunity, the first 1T/cell ROM-based CiM macro has been designed in 28nm CMOS, showing 25.6x higher density than the 6T SRAM-CiM counterpart.

Second, to compensate the limitations by the fact that the weights stored in the ROM-CiM macro could not be updated, this paper proposes an effective convolution weight fine-tune technique called Residual Branch (ReBranch). ReBranch is capable of transferring the neural network from a pretrained generalized model, e.g. VGG-8 and ResNet-18 on CIFAR-100, to support many other specific tasks, e.g. CIFAR-10/Fashion-MNIST/Caltech101, with negligible accuracy loss.

This work sheds light on a new promising CiM category. This early exploration has unveiled the potential for higher capacity and energy efficiency beyond SRAM-based CiM, with further opportunities of circuit, architecture, software, and cross-layer co-optimizations in the future.

Next, section 2 provides the background of ROM and residual block. Section 3 presents the proposed ROM-CiM concept, macro, the ReBranch-assisted architecture, and the YOLoC framework for object detection. Section 4 provides the benchmarking results.

# 2 Background

## 2.1 Read-Only Memory (ROM)

Figure 2 shows the comparison between a practical read-only memory (ROM) and a 6T SRAM, in which one possible 1T/cell ROM implementation is illustrated. Because of the non-volatile nature, ROM also benefits from low standby power and high reliability of read and write disturbance immunity. Also, it is apparent that ROM could be much denser than SRAM.

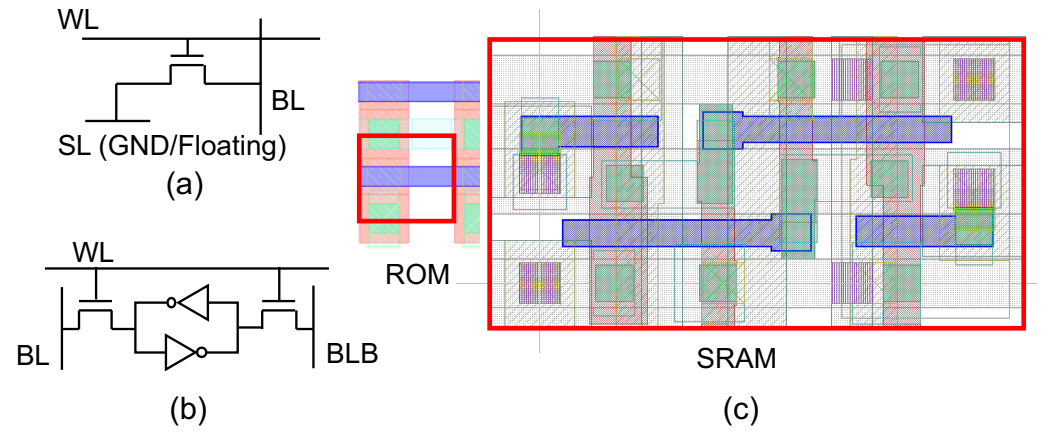

**Figure 2: Illustration of (a) ROM, (b) SRAM, and (c) layout.**

The biggest challenge of replacing SRAM with ROM is the lack of flexibility due to fixed data. Therefore, ROM is usually for scenarios with no data update. If large-scale neural networks could be deployed on ROM-based accelerators, the problems caused by the low density of SRAM will be greatly alleviated.

## 2.2 Residual Block

Residual block is a simple, efficient, and common structure in neural networks to prevent the gradient disappearing, especially for large-scale DNN with more network layers [11]. Its diagram in ResNet is shown in Figure 3(a). The direct connection between the input and the output passes gradient in backpropagation to avoid the gradient disappearing. The convolution layers, i.e. Conv2d in Figure 3(a), fit the residuals.

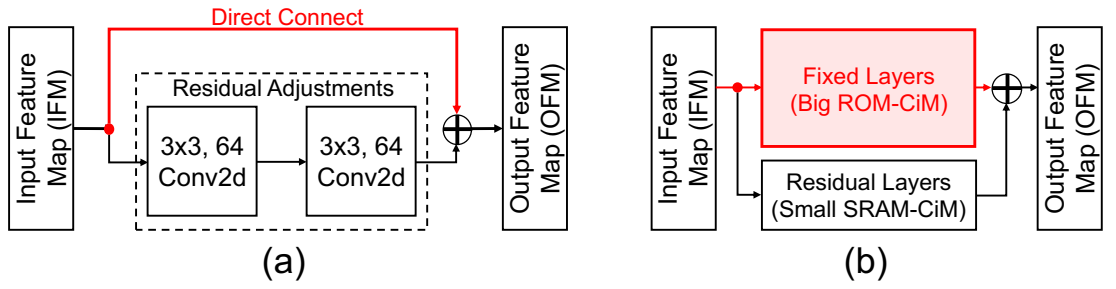

**Figure 3: Diagram of (a) residual block in ResNet and (b) motivated residual branch.**

We introduce residual block as it motivates the ReBranch method in this work to overcome the flexibility challenge in ROM CiM. The key motivation is an assisting SRAM-based convolution block (for additional residual adjustments) could be added in parallel to the fixed ROM-based convolution block (for the pretrained parameters), as shown in Figure 3(b). Therefore, we can learn the additional residuals caused by transfer learning and compute in a fashion of "big ROM + small SRAM". ReBranch will be described in detail in section 3.

## 2.3 Related Works

*Model compression.* To relieve the memory access bottleneck (also computing costs), model compression efforts have been proposed. One category of method is weight pruning [12], post quantization and training-aware quantization [13]. However, ultra-scaled networks below 8-bit quantization, such as TNN [14] and BNN [15], are still difficult to implement on modern networks like ResNet [11] and MobileNet [16]. Another model compression method of using much smaller networks could cause accuracy degradation, especially for complex tasks such as object detection [17]. In addition, chiplet integration in Figure 1(b) is another promising approach towards a larger-scale computing platform, but it still faces the challenge of high inter-chip communication costs and it does not reduce the total chip area cost for the task.

*Beyond-CMOS CiM.* Emerging beyond-CMOS devices, such as RRAM [6], MRAM [16], FeFET [7], etc., provide both non-volatility and higher density for CiM when normalized to the technology feature size. However, due to their early infancy or built-in device operation mechanism limits, there are still major challenges to overcome, involving the device variations, integration capacity, reliability, or computing accuracy.

*CMOS CiM.* Both SRAM-based CiM and eDRAM-based CiM have been reported [3][4][5]. SRAM CiM has been investigated more thoroughly, and recent advance makes eDRAM CiM intriguing because of the higher density. However, eDRAM CiM faces accuracy and refresh challenges due to leakage. Generally, although eDRAM CiM has gained higher density, CMOS-based solutions are currently limited to light computing tasks, or are in need of frequent data access to the external DRAM. This work extends the CMOS-based CiM to the ROM region, and provides opportunities beyond existing SRAM and eDRAM approaches.

# 3 Proposed YOLoC Architecture

## 3.1 ROM-CiM Circuits

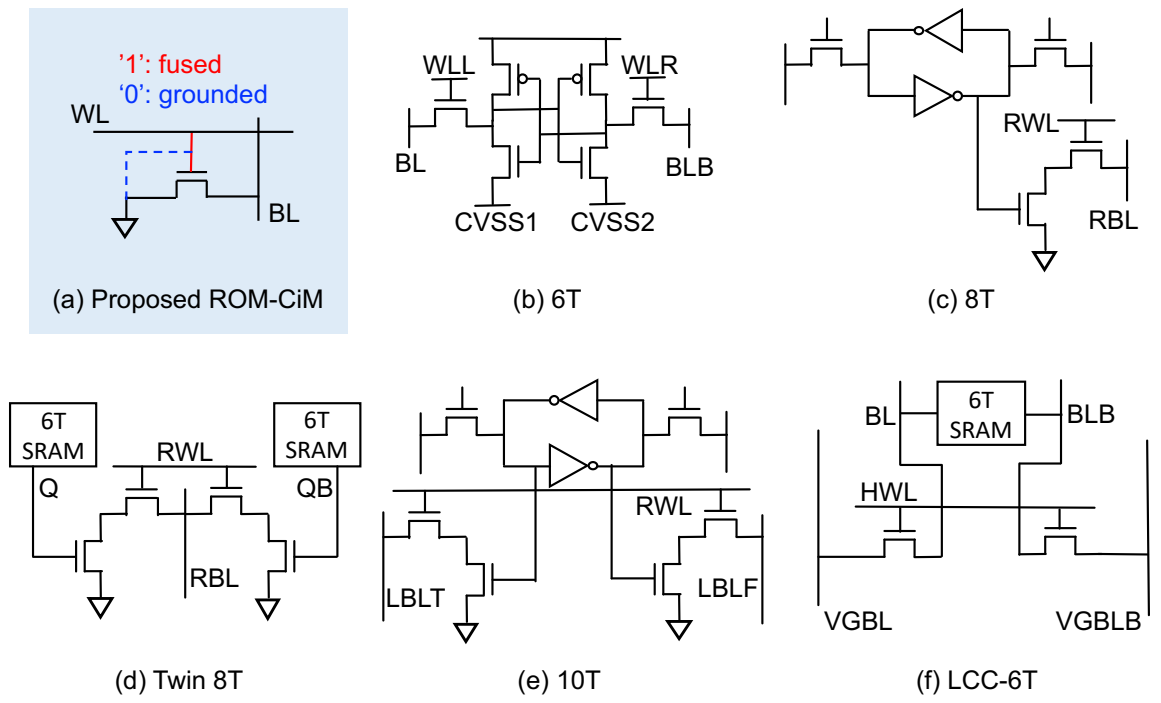

**Figure 4: Proposed 1T/cell ROM-based CiM cell in (a), and some existing SRAM-CiM cells in (b-f) [3][4][19][20][21].**

Figure 4 shows the proposed 1T/cell ROM-based CiM cell circuit, in comparison with some existing SRAM CiM cells [3][4][19][20][21]. Figure 4(a) is one proposed design example. It stores '0/1' by physically connecting the access transistor gate to word line (WL) or a fixed voltage (typically ground). The cell density of the proposed ROM-CiM cell in Figure 4(a) is much higher than the SRAM-CiM cells (14.5-29.5x in our samples).

The computing functionality is carried out between the input WL and the stored '0/1' weight. Only when both the input is high and the weight is physically connected to WL, BL will be connected to the ground. Otherwise, BL will be left floating.

Figure 5 presents the proposed ROM-CiM macro as an example with 128 x 256 cells, 16 column-sharing 5-bit ADCs, input serial-bit drivers, and other peripheral circuits. Cells of the same column or row are connected by sharing BL or WL. The BLs are initially pre-charged before the computing. In the computing, the serial activation input bits are applied to the WLs in a one-cycle-one-bit fashion, in which the number of unary pulses represents the activation amplitude (0, 1, 2, or 3 pulses applied to each WL for a 2-bit activation input). The input activation encoding method using the pulse width may also be used with a different speed-accuracy trade-off.

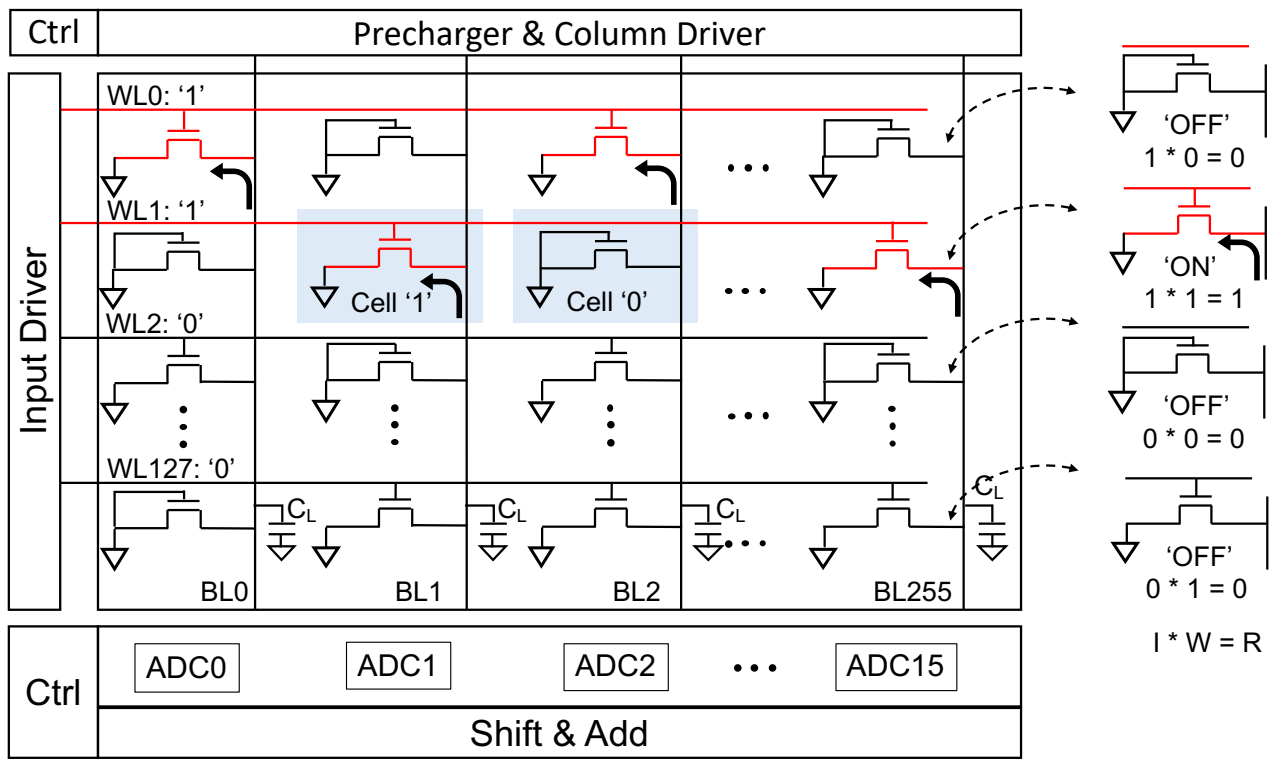

**Figure 5: Proposed ROM-based CiM macro structure, shown with a 128x256 array and 16 ADCs as an example.**

When multiple rows of cells are activated at the same time, the results are summed up over the bitline (BL). Specifically, the charge on the bit line is released to ground according to the number of turned-on cells in the column. The bitline voltage is then sensed by the subsequent ADC to digitize the MAC result. As ADCs take more than one column space, ADC sharing could be applied. In addition, multiple subarrays in the chip could be activated simultaneously to compute with high parallelism.

## 3.2 Residual Branch (ReBranch)

*Goal.* The direct replacement of SRAM CiM with ROM-based scheme has its own limitation in the weight update flexibility, as mentioned above. The goal of the proposed ReBranch is to provide flexibility by introducing a portion of SRAM CiM while still maintaining the feature of high density with ROM CiM and accuracy.

*Option One: ROM-CiM-based One-Shot Learning (ROSL).* As shown in Figure 6(a), the one-shot classification architecture by meta-learning [22], could be adopted by replacing the feature extractor layers with the proposed ROM-CiM, and keeping the feature mapping classifier implementation in an SRAM-based TCAM distance calculator. With the growing maturity of meta-learning, the strength of this scheme is becoming more feasible: when the training set is small, the classification accuracy is higher than other training methods as it could prevent over-fitting. The weakness may also be critical: (i) no accuracy advantage compared with other training methods when the training set is large (one-shot learning-oriented issue) [22], and (ii) difficulty of transitioning to a different task domain, e.g. from character recognition to traffic analysis, with a fixed feature extractor (ROM-oriented issue).

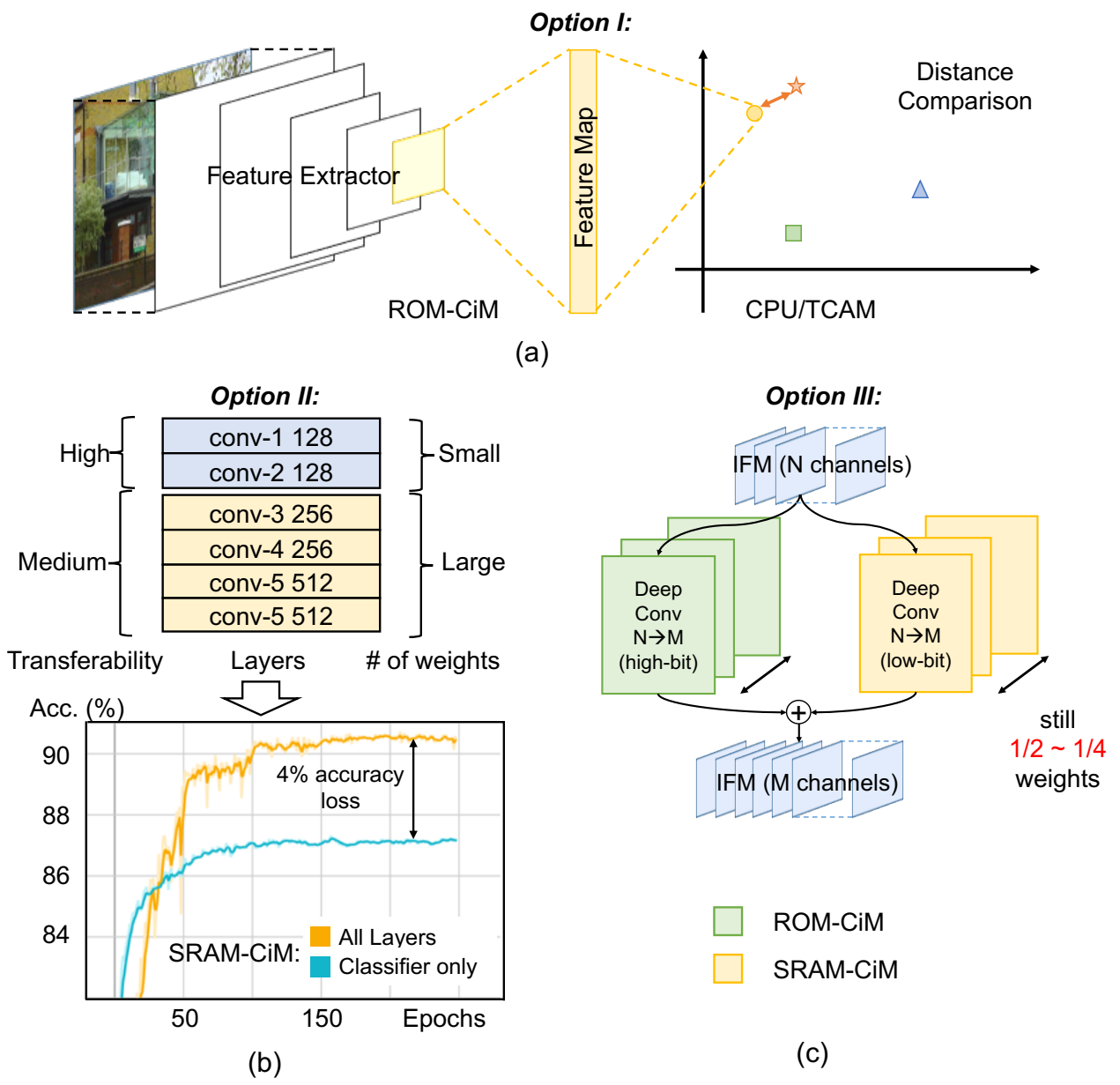

**Figure 6: Options to provide model flexibility to ROM-CiM: (a) Option I: ROM-CiM-based one-shot learning, (b) Option II: alternative transfer learning, (c) Option III: SRAM-assisted parallel weight decoration.**

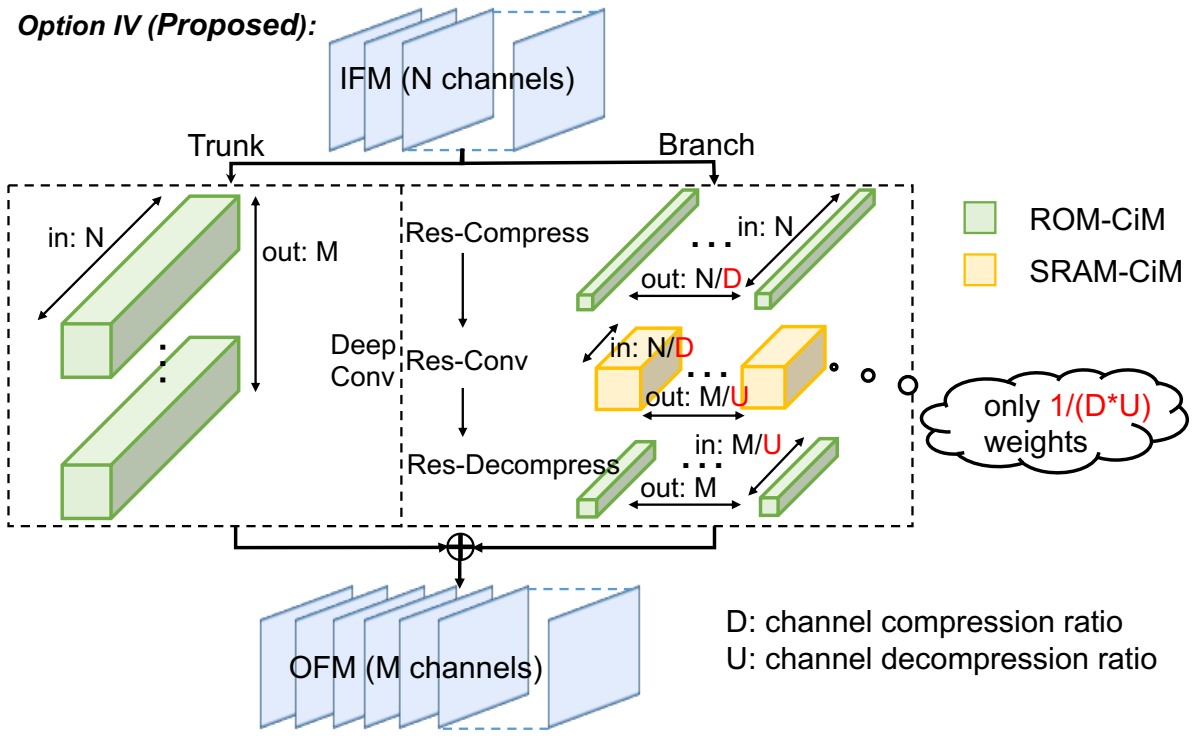

**Figure 7: Proposed ReBranch structure.**

*Option Two: Alternative Transfer Learning (ATL).* As shown in Figure 6(b), when transferring a pretrained model to a new target dataset, the weights of some layers could be fixed and implemented in ROM CiM, while others could be alternatively implemented in SRAM CiM. Practically, the first few layers have higher transferability and are less likely to be modified during transfer learning. However, it could be difficult to deploy many ROM-CiM layers due to transferability decay when going deep, as shown by the experiments in Figure 6(b).

*Option Three: SRAM-Assisted Parallel Weight Decoration (SPWD).* As shown in Figure 6(c), a low-weight-quantization SRAM CiM branch could be deployed in parallel with the ROM CiM branch so as to modify the changes after transfer learning. Practically, to maintain the accuracy, a typical SRAM CiM quantization of 2 is needed to decorate the 8-bit ROM CiM quantization, leading to a maximum area-saving of 4x.

*Option Four: Proposed Residual Branch (ReBranch).* To overcome the weaknesses of the prior three options, Figure 7 shows the ReBranch structure motivated by the residual block in ResNet [11]. ReBranch consists of two parallel blocks: the *trunk* and the *branch*. The *trunk* is a deep convolution layer group with fixed parameters. The *branch* consists of two parameter-fixed residual-(de)compression layers and one trainable residual-convolution layer group. The residual compression layer and the residual decompression layer are used to transform the number of channels of the feature map. With this structure, all the trunk layers and Res-(De)compress layers could be deployed in high-density ROM-CiM, while the number of parameters in Res-Conv deployed in SRAM-CiM is much smaller. Practically, the branch convolution layer could be 16x smaller than the trunk layer.

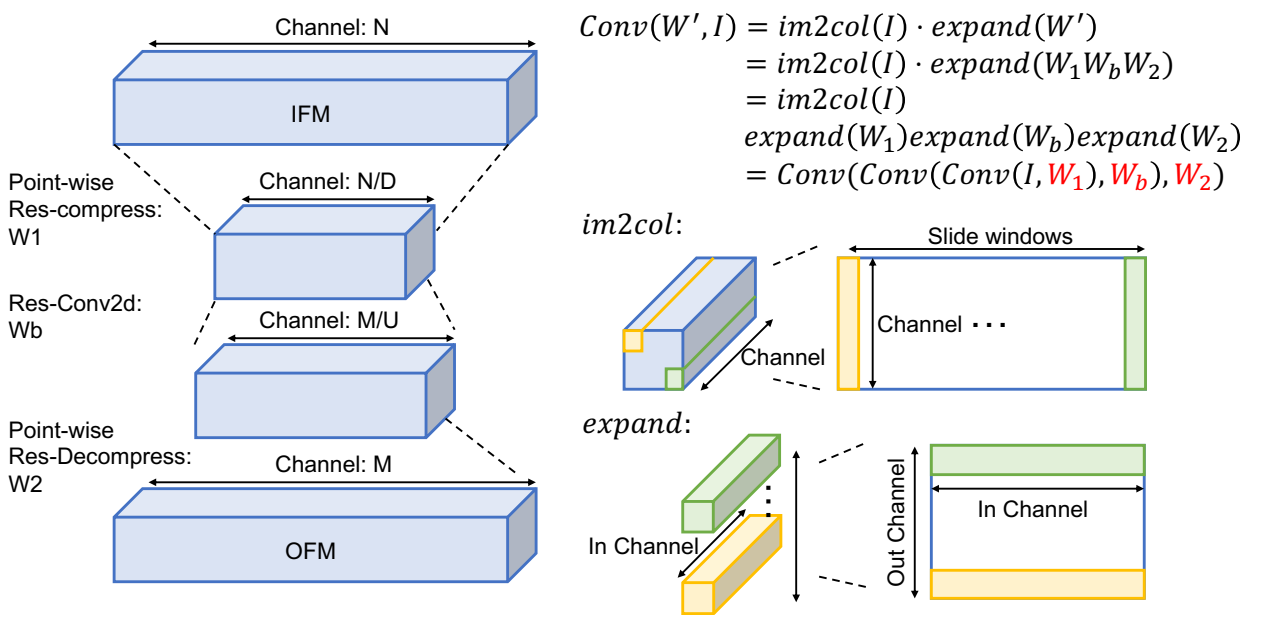

**Figure 8: Diagram of point-wise convolution for channel compression and decompression.**

We adopt point-wise convolution [23] to compress and decompress the channels of the feature map. This kind of *branch* design can be equivalent to the same size convolution layer as the *trunk*, as shown in Figure 8. In principle, the *trunk* layer parameters in ROM-CiM could be adjusted to a certain extent by trainable SRAM-CiM. When transferring to a new dataset, the *branch* essentially learns the residual of the trunk to ensure high classification accuracy.

In the proposed ReBranch, the optimization goal is to achieve minimum area occupation by designing proper Res-(De)Compression layers, which leads to the reduction of the number of channels used in Res-Conv. Experimental results in section 4 will further reveal the optimization of this structure.

## 3.3 Proposed YOLoC Computing Framework

Taking advantages of the ReBranch-assisted ROM-CiM, we propose the YOLoC framework, as shown in Figure 9. It is composed of ROM-CiM, SRAM-CiM, SRAM cache, and controller. Corresponding to the logic flow, ROM-CiM is responsible for the inference of the backbone network, which accounts for most of the parameters and calculations. SRAM-CiM is used for ROM-CiM parameter fine-tune by ReBranch and feature prediction. SRAM cache is used to store the intermediate data, including non-CiM computing data of activation function and pooling. The controller is used to schedule data and non-CiM computing.

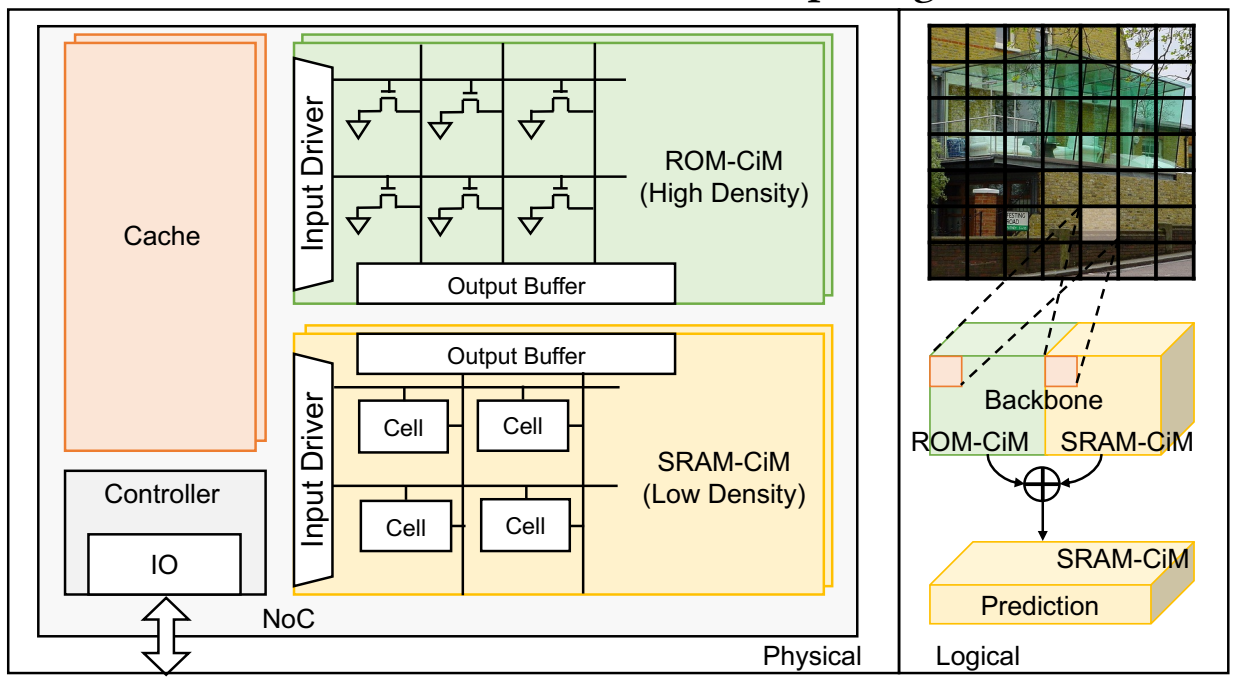

**Figure 9: Proposed architecture of YOLoC.**

In YOLoC, only a small part of weights needs to be loaded from off-chip DRAM to on-chip SRAM-CiM at power-on, which will reduce a lot of overhead in data movement. Over 90% of parameters are stored in the high-density ROM-CiM. In addition, it also provides a chance to greatly reduce the on-chip training overhead, especially when performing on-chip large-scale neural networks training [8] in SRAM-CIM.

# 4 Experimental Result

## 4.1 Environment Setup

Two image classifier models, VGG-8 and ResNet-18, and the object detection model YOLO (DarkNet-19 backbone) are used in our benchmarks. SRAM-CiM and ROM-CiM macro parameters are all obtained from parasitic extraction and SPICE simulation in 28nm CMOS. The system-level simulation for accuracy, area, latency, and energy per inference is based on our custom workflow simulator by PyTorch.

## 4.2 ReBranch Generalization

Figure 10 shows the ReBranch generalization evaluation results on VGG-8 and ResNet-8 pretrained on CIFAR-100 (C100). Figure 10(a-b) compares accuracy and memory area on different datasets using VGG-8 and ResNet-18 as models. ReBranch saves area by 10x than the all-SRAM-CiM baseline with only <0.4% accuracy loss in image classification. In practice, the model deployed in ROM should be trained on a broader dataset to accommodate a broader migration of applications.

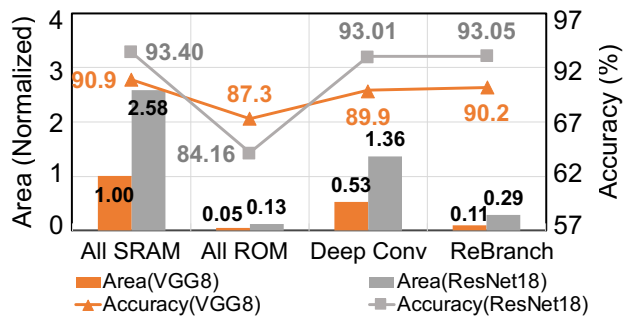

| Method | ROM | SRAM |
|---|---|---|
| All SRAM [3] | - | feature extractor |
| All ROM (Option II) | feature extractor | - |
| Deep Conv (Option II) | except last conv | last conv |
| ReBranch (Option IV) | feature extractor | res-conv |

(a)

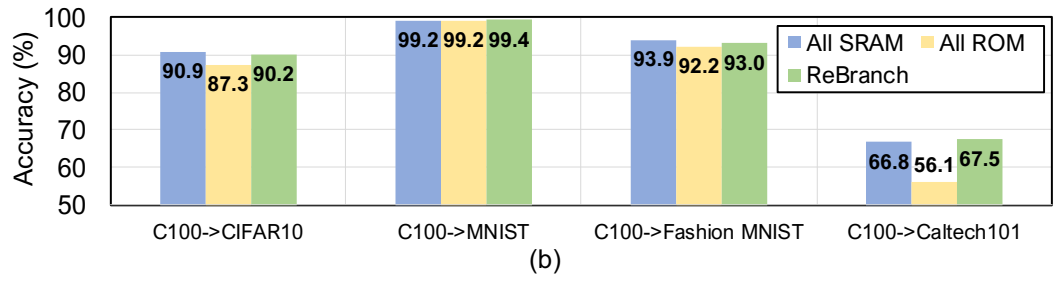

(b)

**Figure 10: Generalization analysis of residual branch.**

To achieve optimized area savings and generalization trade-off, the ReBranch hyperparameters, i.e. the compression ratio $D$ and the decompression ratio $U$ in Figure 7 are investigated. As shown in Figure 11(a), with D=U=4, the maximum accuracy of 93.1% and 90.2% could be achieved for ResNet-18 and VGG-8, respectively. Note that the $D^*U$ is the overall parameters compression ratio, which determines the trade-off between the area saving and the model flexibility. Figure 11(b) shows that 16x compression is a reasonable choice in VGGNet and ResNet for good accuracy and overall area savings.

Figure 12 compares the chip area and mean average precision (mAP) in the PASCAL VOC dataset. For chip area, the proposed YOLoC outperforms SRAM-CiM using YOLO and Tiny-YOLO (a smaller net) by 9.7x and 2.4x, respectively. YOLoC also shows significant mAP improvement over *Option 2: Alternative Transfer Learning*, and almost no mAP loss (-0.5% ~ 0.2%) compared with the SRAM-CiM baseline.

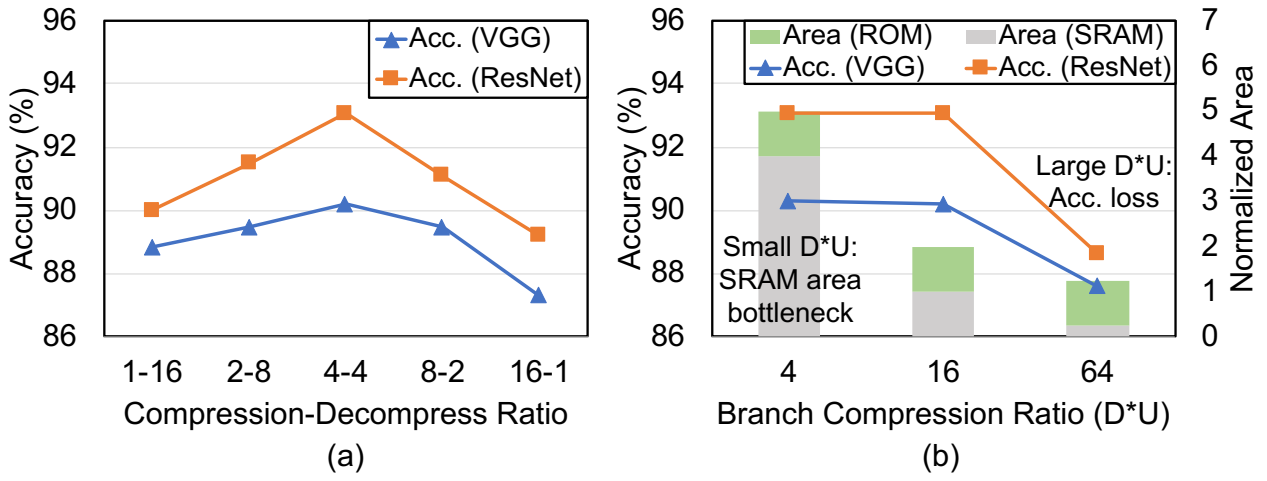

**Figure 11: Result of parameters analysis in Residual Branch .**

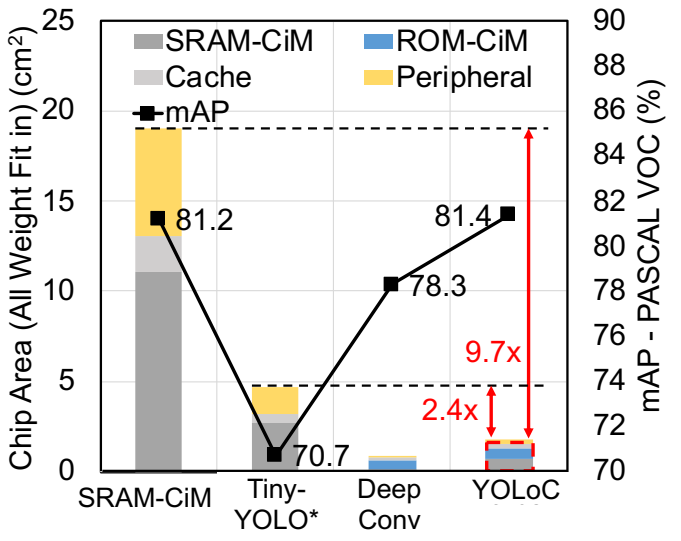

| mAP | COCO → Pedestrian Detection | COCO → Traffic Detection | COCO → PASCAL VOC |
|---|---|---|---|
| All Layers Trainable (SRAM-CiM [3]) | 83.6% | 87.0% | 81.2% |
| Only Prediction Trainable (Option II) | 81.2% | 85.3% | 78.3% |
| **Proposed Residual Branch (Option IV)** | **83.1%** | **87.1%** | **81.4%** |

* Tiny-YOLO: a smaller backbone in the same framework (all layers trainable).

**Figure 12: mAP and memory area in different methods.**

## 4.3 System Evaluation

*4.3.1 Macro evaluation.* The proposed ROM-CiM macro in Figure 5 is evaluated. Table I summarizes the specifications of the proposed ROM-based CiM macro. *Note* that the weight reload overhead is not considered at this macro level, but at the system level (section 4.3.2). The proposed ROM-based CiM cell has achieved a density record of 0.014μm²/bit among existing CMOS-based CiM cells. It is 16x smaller than a compact-rule 6T SRAM in the same process and 18.5x smaller than the recent SRAM-CiM cell in [3]. Actually, it is even denser than the commercial SRAM at the 5-7nm node. The peripheral of the macro is also smaller than the SRAM-CiM counterpart with a simplified read and write IO interface. Using the computing peripheral circuits from [3], ROM-CiM achieves a record-high density of 5Mb/mm², which is 19x larger than SRAM-CiM in the same 28nm process.

Furthermore, the flexibility of sensing the MAC result by digitizing the remnant charge of a pre-charged bitline capacitor is the same as [3] and other recent SRAM CiM works. This makes SRAM CiM circuit optimizations potentially applicable to the proposed ROM CiM macro, too, provided that the increased density is considered. For example, the trade-off between the number of ADCs and simultaneously activated rows. This could be explored in future works.

**Table I: ROM-CiM macro specification summary**

| | |
|---|---|
| Process | 28nm CMOS |
| Macro size | 1.2 Mb |
| Macro area | 0.24 mm² |
| Macro density | 5 Mb/mm² (25.6x) |
| Cell area | 0.014 μm² |
| Input x weight | 8-bit x 8-bit |
| Inference time | 8.9 ns |
| Operation number | 256 |
| Throughput | 28.8 GOPS |
| Macro area efficiency | 119.4 GOPS/mm² |
| MAC energy efficiency | 11.5* TOPS/W |
| Standby power | 0 (non-volatile) |

*: Data estimated using peripherals from [3] (error range: <7%).

*4.3.2 System evaluation results.* Based on the macro specifications, the system evaluation of YOLoC is carried out. The read/write energy and latency of SRAM buffer and DRAM are obtained by CACTI [24]. The weight mapping scheme is optimized in a way of storing the weights of different layers to the same sub-array, so as to achieve high ADC utilization and thus reduced latency. Note that the ROM-CiM is more compact than SRAM-CiM with a simplified R/W interface, the iso-area comparison is achieved by adopting more sub-arrays for ROM-CiM in the evaluation.

Figure 13 shows the three system evaluation configurations. The proposed YOLoC uses SRAM-CiM to assist ROM-CiM transfer to various tasks. The iso-area single-chip SRAM-CiM method requires DRAM to store the additional weights that on-chip SRAM-CiM cannot store. The chiplet method makes use of multiple SRAM-CiM chips to store all the model parameters so that no DRAM is needed. For the chiplet method, intermediate data need to be transferred via chiplet interconnection.

Figure 14 shows the chip-level comparisons with recent CMOS-based CiM works, regarding the area efficiency, energy efficiency, and the breakdown of energy and area. Compared with the iso-area single SRAM-CiM chip, the energy efficiency improvement with ResNet-18 and Tiny-YOLO is 4.8x and 10.2x, respectively. For larger and more complex networks that cannot be crammed onto a single chip, DRAM data access of SRAM-CiM brings significant overheads. In this case, YOLoC achieves 14.8x energy efficiency improvement over SRAM-CiM in the YOLO (Darknet-19 backbone) model. The existence of a residual branch introduces little latency overhead. On YOLO, the latency overhead is only 8%. Compared with the SRAM-CiM chiplet solution, the proposed YOLoC achieves ~2% energy efficiency improvement with less data movement overhead, plus significant total chip area savings of ~10x.

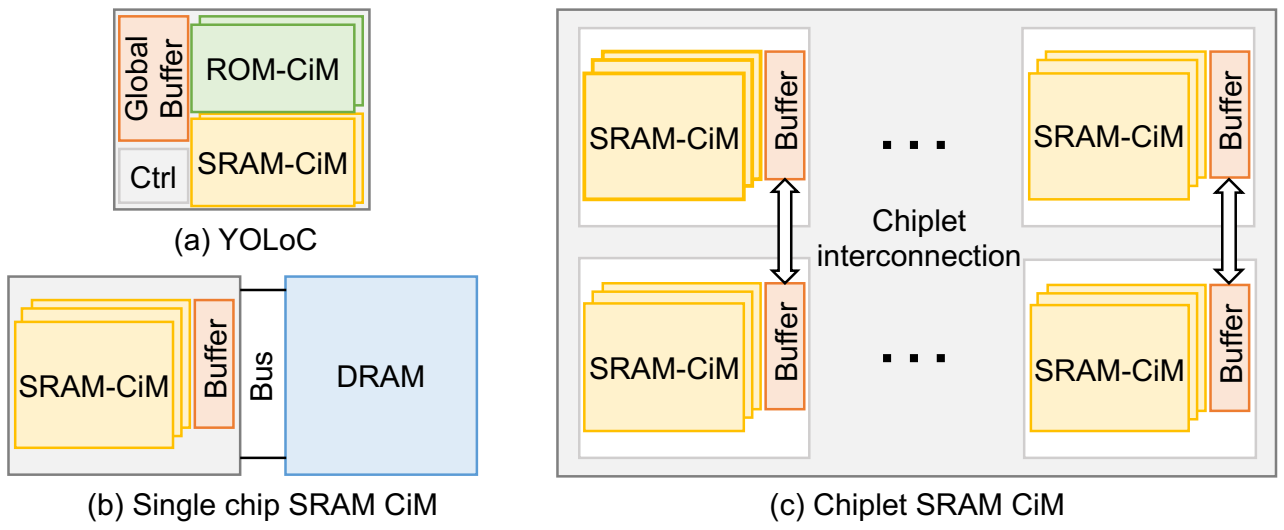

**Figure 13: System configuration of (a) proposed YOLoC, (b) single-chip SRAM-CiM, and (c) SRAM-CiM chiplets. Chiplet interconnection parameters are from SIMBA[25]**

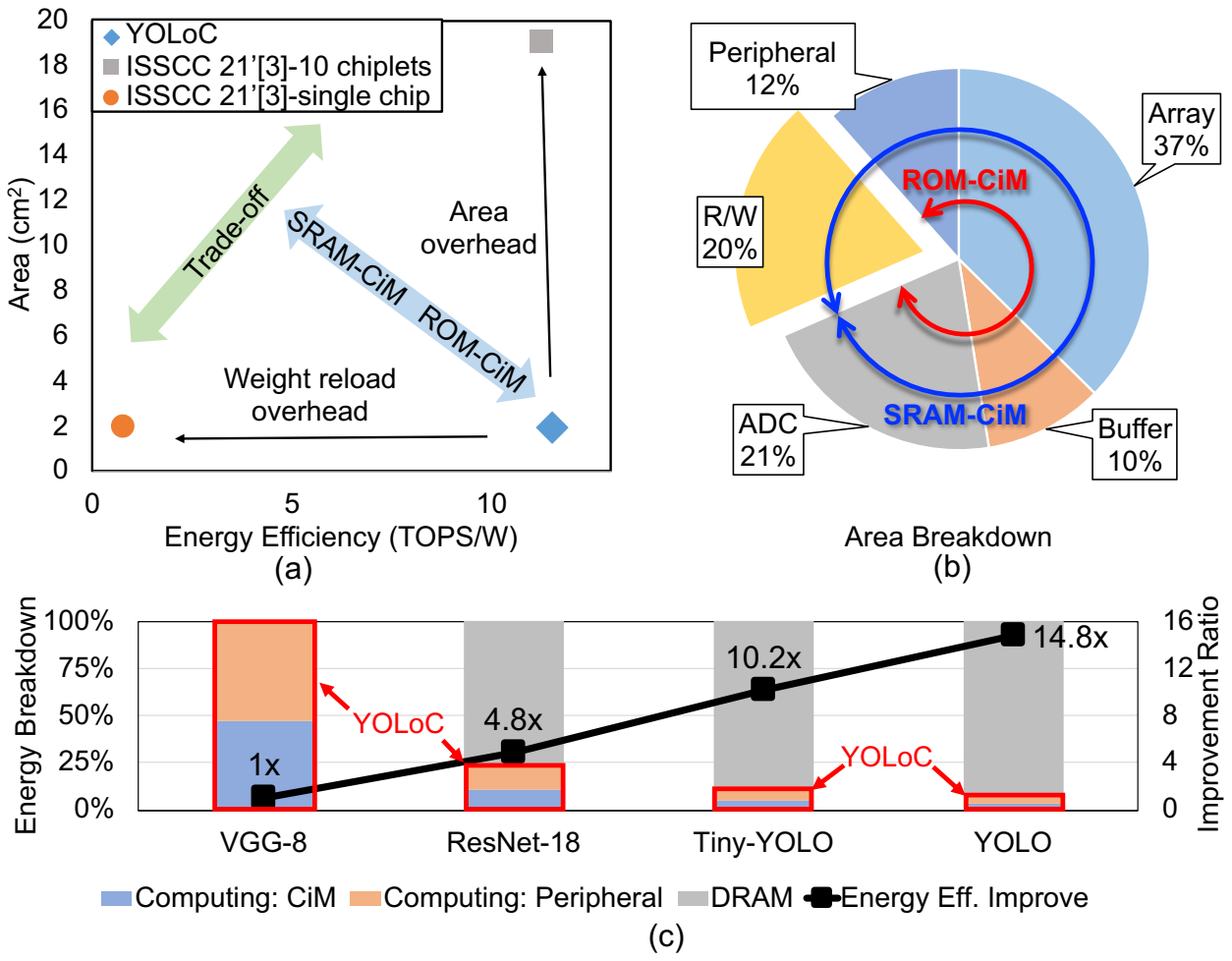

**Figure 14: Chip-level comparisons with recent SRAM-based CiM works: (a) overview, (b) area breakdown, (c) energy breakdown of SRAM-CiM with different NN models.**

*4.3.3 Perspectives.* Ultra-high CiM cell density is the most significant progress made by the proposed ROM-CiM. Because the pure SRAM-CiM deployment of a large-scale model on one single chip is limited by the high DRAM access overhead, the actual performance may degrade dramatically, despite data reuse. Ping-Pong and pipelining techniques can relieve the latency issue, but little could be done to the energy overhead while designing an SRAM-CiM macro. ROM-CiM shows the opportunity of dealing with this situation. Future works that thoroughly exploit the ROM-CiM design space and cross-layer co-optimizations (including ROM-CiM chiplets) are promising.

## 5 Conclusions

In this paper, an ultra-high-density versatile ROM-based CiM method and the large-scale neural network framework YOLoC assisted by ReBranch have been proposed to solve the CMOS CiM density bottleneck. The presented ROM-based CiM array achieves a record-high CMOS CiM array density. The proposed ReBranch is capable of transferring the neural network in ROM-CiM from a pretrained model to various tasks. Moreover, the proposed YOLoC framework is capable of deploying a complete YOLO model onto a single CMOS-based chip without weight reloading from DRAM. Finally, the evaluations of the proposed techniques prove high generalization, area efficiency, and energy efficiency, suggesting a new paradigm for data-intensive neural network acceleration.

## 6 Acknowledgment

This work is supported in part by National Key R&D Program of China (2019YFA0706100), and in part by NSFC (U21B2030, 61874066, 61720106013, 61934005).